\documentclass[prl,amssymb,intlimits,centertags,amsmath,twocolumn
               ,showpacs,nofootinbib,floatfix]{revtex4}
\usepackage{graphicx}
\usepackage{dcolumn} % to align columns in tables
\usepackage{bm} % for bold face in math environment
\allowdisplaybreaks
\begin{document}

 \title{The Extreme Distortion of Black Holes due to Matter}
 
 \author{David Petroff}
 \email{D.Petroff@tpi.uni-jena.de}
 \affiliation{Theoretisch-Physikalisches Institut, University of Jena,
           Max-Wien-Platz 1, 07743 Jena, Germany}
 
 \author{Marcus Ansorg}
 \email{Marcus.Ansorg@aei.mpg.de}
 \affiliation{Max-Planck-Institut f\"ur Gravitationsphysik,
              Albert-Einstein-Institut, 14476 Golm, Germany}

 \date{\today}

 %\selectlanguage{english}
 \begin{abstract}
  A highly accurate computer program is used to study axially symmetric
  and stationary spacetimes containing a Black Hole surrounded by a ring
  of matter. It is shown that the matter ring affects the properties of
  the Black Hole drastically. In particular, the absolute value of the ratio
  of the Black Hole's angular momentum to the square of its mass not only
  exceeds one, but can be greater than ten thousand 
  ($|J_\text c|/M_\text c^2 > 10^4$). Indeed, the numerical evidence
  suggests that this quantity is unbounded.
 \end{abstract}

 \pacs{04.70.Bw, 04.40.-b, 04.25.Dm, 04.70.-s \qquad preprint number: AEI-2005-176}
 
 \maketitle
 
 One of the great contributions to general relativity was Roy Kerr's
 presentation in 1963 \cite{Kerr63} of an axially symmetric, stationary
 solution to Einstein's vacuum equations that turned out to describe a rotating
 Black Hole -- the Kerr Black Hole. The fact that this simple solution
 is fully described by specifying merely two parameters, the fact that there
 exist proofs (see for example \cite{Heusler96}) showing it to be the only
 stationary
 (non-charged) Black Hole and the fact that recent observations show Black
 Holes to be important and ubiquitous objects in the cosmos, make the Kerr
 solution all the more important. The Black Holes known to exist in the centres
 of many galaxies play an important role in present day astronomy and compact
 binaries  containing at least one Black Hole are expected to be important
 sources for the burgeoning field of gravitational wave astronomy.
 Such objects are, however, anything but isolated, whence the Kerr solution is
 not always going to be appropriate to model the corresponding Black Hole.
 
 In this letter we consider a stationary and axisymmetric spacetime containing
 a Black Hole surrounded by a ring of matter. We write the line element in
 Lewis-Papapetrou coordinates
 \[ ds^2 =  e^{2\mu}(d\varrho^2+d\zeta^2)
           + \varrho^2 B^2 e^{-2\nu}\left(d\varphi - \omega dt \right)^2
           - e^{2\nu}dt^2, \]
 where the four metric functions $\mu$, $B$, $\nu$ and $\omega$ depend only
 on $\varrho$ and $\zeta$. We make use of a coordinate freedom to choose the
 horizon to be a sphere in these coordinates and ensure that this surface is
 indeed an event horizon by requiring that
 \[ e^{\nu} = 0,\quad B=0 \quad \text{and} \quad 
      \omega=\text{constant}=:\Omega_\text c\]
 hold there \cite{Bardeen73}. The functions $e^{\nu}$ and $B$ tend to
 zero on the horizon in such a way that
 \[ e^u := e^\nu /B\]
 remains finite and non-zero. Einstein's equations along with these boundary
 conditions, the correct asymptotic behaviour, transition conditions and
 regularity make up a free-boundary problem (the surface of the ring is
 unknown {\em a priori}), which we solve using a
 multi-domain pseudo-spectral method described in detail in \cite{AP05}.
 After specifying the equation of state used to describe the perfect fluid
 ring, the solution depends on four parameters. In this letter we choose the
 very simple model of a uniformly rotating ring of constant density.
 
 It is well known that for the Kerr solution, the dimensionless quantity%
 \footnote{We choose natural units in which the speed of light $c$ and
 gravitational constant $G$ are both equal to one.}
 $|J|/M^2$ describing the ratio of the angular momentum to the square of the
 mass of the Black Hole can range from zero in the static limit to one for
 the extreme Kerr Black Hole. Recently, steps have been taken toward showing that
 this quantity must be less than one for all axially symmetric,
 asymptotically flat, complete pure vacuum data \cite{Dain05, Dain05b, Dain05c}. In
 the case being considered here, it is possible
 to define the mass and angular momentum of the Black Hole and the ring
 separately. It follows from the field equations that the total angular
 momentum of the system can be represented as a surface integral over the
 horizon of the Black Hole plus a volume integral over the ring. The integrals
 are the same as those one finds for the case of a lone Black Hole or ring
 so that one can define this local object as `belonging' to the body in
 question. The situation for the mass of the two objects is very similar (see
 \cite{Bardeen73} for more details).
 It was shown in \cite{AP05} that the inequality
 $|J_\text c|/M_\text c^2 \le 1$, where the subscript `c' indicates that the
 objects refer to the Black Hole (i.e.\ the central object), no longer holds
 for a Black Hole surrounded
 by a ring of matter. There, an exemplary configuration in which  $|J_\text c|/
 M_\text c^2 = 20/19$ was calculated to high accuracy. We now show that this
 quantity can become dramatically larger than one.
 
 We consider a sequence of configurations for which the angular velocity
 of the horizon with respect to infinity is taken to be $\Omega_\text c=0$,
 the ratio of the inner to
 outer coordinate radius of the ring is held constant at a value of
 $\varrho_\text i/\varrho_\text o=0.7$ and a geometric parameter measuring
 the `distance' of the outer edge of the ring to a mass-shedding limit
 is $\beta_\text o=0.1$ ($\beta_\text o=0$ corresponds to the outer 
 mass-shedding limit)%
 \footnote{The definition of $\beta_\text o$ is 
           $\frac{2\varrho_\text o}{\varrho_\text o -\varrho_\text i}
             \left. \frac{d(\zeta_\text B^2)}{d(\varrho^2)}
             \right|_{\varrho=\varrho_\text o}$, where
           $\zeta_\text B = \zeta_\text B(\varrho)$ is a parametric
           representation of the surface of the ring.}.
 We can then choose the ratio of the Black Hole mass to that of the ring
 $M_\text c / M_\text r$ to parameterize the one dimensional sequence that
 results. The choice $\beta_\text o=0.1$ ensures that the outer edge of the
 ring always remains close to the mass-shedding limit, which, in turn means
 that the influence of the Black Hole on the ring is never negligible, 
 even in the limit $M_\text c / M_\text r \to 0$. This can be seen by
 examining the behaviour of a sequence of homogeneous rings
 without a central body, where $\varrho_\text i/\varrho_\text o=0.7$ is
 held constant. For such a sequence, the outer mass-shedding parameter ranges
 from $\beta_\text o=0.832$ in the Newtonian limit to $\beta_\text o=0.463$ in
 the extreme Kerr limit. This sequence is a vertical line in Fig.~1 of
 \cite{AKM4} and can be seen to remain far away from the mass-shedding curve.
 Hence the choice $\beta_\text o=0.1$ is only possible by virtue of the
 influence of the central Black Hole.

 \begin{figure}
  \includegraphics{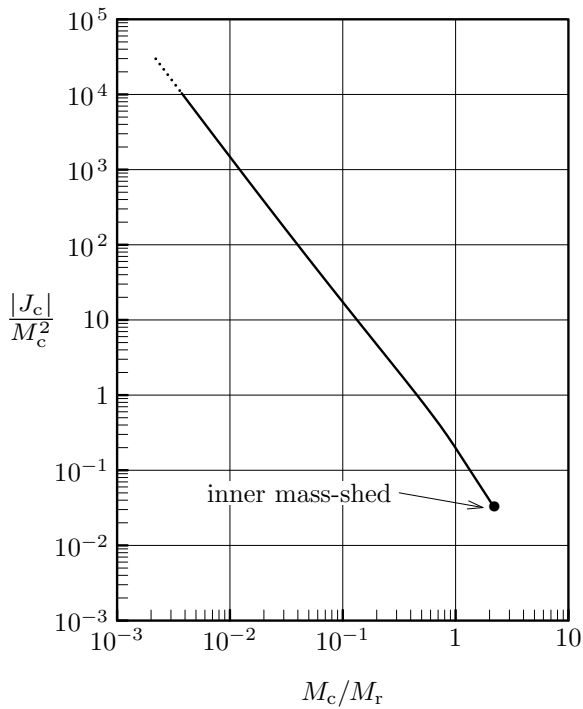}
  \caption{The dimensionless quantity $|J_\text c|/M_\text c^2$ is 
           plotted versus $M_\text c/M_\text r$ for a sequence
           in which $\Omega_\text c=0$, $\varrho_\text i/\varrho_\text o=0.7$
           and $\beta_\text o=0.1$ were held constant. The dotted line
           indicates that the curve continues on further.\label{M2J}}
 \end{figure}

 Fig.~\ref{M2J} reveals the dramatic influence the matter ring can
 have on the Black Hole. Beginning with the right side of the curve,
 we see that for large values of $M_\text c/M_\text r$, $|J_\text c|/
 M_\text c^2$
 grows small, just as expected when nearing a Schwarzschild Black Hole. The
 effect of the ring is no longer so great. As $M_\text c/M_\text r$ increases,
 the Black Hole `pulls' more and more on the ring until an inner mass-shedding
 limit is reached.
  
 \begin{figure}
  \includegraphics{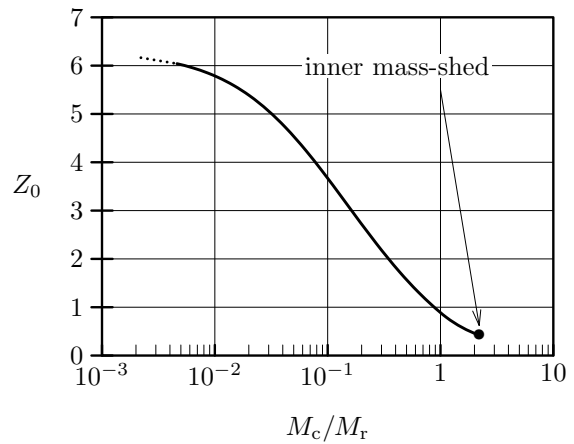}
  \caption{The relative redshift of zero angular momentum photons
           emitted from the surface of the ring and observed at infinity is 
           plotted versus $M_\text c/M_\text r$ for the sequence
           plotted in Fig.~\ref{M2J}. The dotted line
           indicates that the curve continues on further.\label{Z0}}
 \end{figure}
 
 Now turning our attention to the left side of the curve in Fig.~\ref{M2J}, 
 we see that as $M_\text c/M_\text r \to 0$, $|J_\text c|/M_\text c^2$ grows
 dramatically. Indeed, there is no indication that there is any bound
 whatsoever to this quantity. This is one measure of the fact that the
 properties of axially symmetric, stationary Black Holes that we know so
 well from the Kerr solution can be affected drastically by the presence
 of matter. That such an extremely large influence is observed for the
 sequence in Fig.~\ref{M2J}, presumably has to do with the fact that a
 situation was chosen in which the surrounding matter can become
 highly relativistic, as measured for example by the relative redshift $Z_0$ of
 a zero angular momentum photon leaving the surface of the ring and observed
 at infinity. In Fig.~\ref{Z0}, $Z_0$ is plotted versus $M_\text c/M_\text r$.
 
 In Table~\ref{tab} various physical parameters are shown for a configuration
 near the left edge of Fig.~\ref{M2J} for which $|J_\text c|/M_\text c^2$
 has already reached a very large value (all digits listed are valid). 
 The symbols  $R_\text p$, $R_\text e$, $R_\text i$ and $R_\text o$ refer to
 circumferential radii (proper circumference/$2\pi$) for the polar and
 equatorial radii of the horizon and the inner and outer radii of the ring
 respectively. The symbol $\kappa$ refers to surface
 gravity and $A_\text c$ to the surface area of the horizon.
 The subscript `c' refers to the Black Hole (i.e. central body)
 and `r' to the ring. The bar above the symbol indicates that the
 quantity is dimensionless through multiplication with the 
 appropriate power of energy density, which we have chosen to be the scaling
 parameter. The last two rows show that the identities $M_\text i = M_\text a$ and 
 $J_\text i = J_\text a$ are satisfied to very high accuracy.
 Here $M_\text i$ and $J_\text i$ refer to    
 the total mass and angular momentum of the system as calculated via
 a volume integral over the ring plus a surface integral over the
 horizon of the Black Hole whereas $M_\text a$ and $J_\text a$ refer
 to the total mass and angular momentum as read off at infinity from
 the asymptotic behaviour of the metric functions. For more details
 regarding the various quantities discussed here, see \cite{AP05}.
 
 If one compares the Black Hole in the configuration of Table~$\ref{tab}$
 with a Kerr Black Hole of either the same dimensionless parameters
 $\bar{J}_\text c$ and $\kappa M_\text c$
 or  $\bar{J}_\text c$ and $\bar{A}_\text c$, then one finds that the first is
 very close to the extreme Kerr solution ($\kappa M_\text c=0$ marks this
 solution after all), whereas the second is significantly farther away, but
 nonetheless quite close, as witnessed by the value 
 $|J_\text c|/M_\text c ^2=0.9998$ for example.
 In both cases, the difference between the pure vacuum case and the case
 containing the matter ring is dramatic.

\newcolumntype{d}[1]{D{.}{.}{#1}}
\renewcommand{\arraystretch}{1.2}
\begin{table}
  \caption{Physical parameters for a configuration near the left edge
    of Fig.~\ref{M2J} (left column). 
    The other two columns provide comparative values from the Kerr solution
    with the two prescribed parameters shown in bold.
    The last two rows show how
    well the identities $M_\text i = M_\text a$ and 
    $J_\text i = J_\text a$ are fulfilled (see text for a description
    of the various symbols).
     \label{tab}}
  \begin{ruledtabular}
  \begin{tabular}{ld{13}d{8}d{9}}
    &&\multicolumn{1}{c}{Kerr with}&\multicolumn{1}{c}{Kerr with}\\
    & \multicolumn{1}{c}{BH--Ring} 
    &\multicolumn{1}{c}{same $\bar{J}_\text c,\kappa M_\text c$} 
	 &\multicolumn{1}{c}{same $\bar{J}_\text c,\bar{A}_\text c$}
   \\ \hline 
   $|J_\text c|/M_\text c ^2$
 & 3.281340 \times 10^4 & 1- 4.5 \times 10^{-10} & 0.9998 \\
   $\bar{J}_\text c$
 & -5.415253 \times 10^{-2} & \bm{-5}.\bm{42 \times 10^{-2}} 
                            & \bm{-5}.\bm{42 \times 10^{-2}}\\
   $\bar{J}_\text r$
 & 0.4383768 & \text{--} & \text{--} \\
   $\bar{M}_\text c$
 & 0.001284647 & 0.233 & 0.233\\
   $\bar{M}_\text r$
 & 0.6124967 & \text{--} & \text{--}\\
   $\bar{\Omega}_\text c$
 & 0 & -2.15 & -2.11\\
   $\bar{\Omega}_\text r$
 & 0.6148853 & \text{--} & \text{--}\\
    $R_\text p/R_\text e$
 & 0.6160728 & 0.608 & 0.616\\
    $\bar{R}_\text  e$
 & 0.4654637 & 0.465 & 0.465\\
    $\bar{R}_\text  i$
 & 1.251534 & \text{--} & \text{--}\\
    $\bar{R}_\text  o$
 & 1.518647 & \text{--} & \text{--}\\
    $\kappa M_\text c$
 & 1.494494\times 10^{-5} & \bm{1}.\bm{49 \times 10^{-5}} & 0.00951\\
    $\bar{A}_\text c$
 & 1.387660 & 1.36  & \bm{1}.\bm{39} \\ \hline 
   $|\frac{M_\text i - M_\text a}{M_\text a}|$ & 2.6 \times 10^{-9} 
	       & \text{--} & \text{--} \\ 
   $|\frac{J_\text i - J_\text a}{J_\text a}|$ & 3.0 \times 10^{-9}
	 & \text{--} & \text{--} \\
  \end{tabular}
  \end{ruledtabular}
 \end{table}

 It will be both interesting and important to study the stability of systems
 consisting of Black Holes surrounded by rings. If it turns out that they
 can be expected to exist for astrophysically relevant periods of time, then
 we are going to have to rethink what conclusions can be drawn from the
 observation of a densely filled region of space. Not only could normal matter
 make a significant contribution to the mass of such a region, but the
 properties of the central Black Hole (if one is expected to exist) are 
 not necessarily going to be akin to those we know from the Kerr solution or
 indeed from a perturbation to the Kerr solution.

% \begin{acknowledgments}
  \paragraph{Acknowledgments}
  We are grateful to R.\ Meinel for many interesting and helpful discussions.
  This work was supported in part by the Deutsche Forschungsgemeinschaft (DFG)
  through the SFB/TR7 ``Gravitationswellenastronomie''. 
% \end{acknowledgments}
  
 \bibliography{Reflink}
  
\end{document}